\documentclass[aps,pre,showpacs,twocolumn,groupedaddress,amssymb]{revtex4}

\usepackage{graphicx}
\usepackage{dcolumn}
\usepackage{bm}

\begin{document}

\title{Hard x-ray or gamma ray laser by a dense electron beam}

\author{S. Son}
\affiliation{18 Caleb Lane, Princeton, NJ 08540}
\author{Sung Joon Moon}
\affiliation{28 Benjamin Rush Ln., Princeton, NJ 08540}
\date{\today}

\begin{abstract}
A coherent x-ray or gamma ray can be created from a dense electron beam
propagating through an intense laser undulator.
It is analyzed by using the Landau damping theory which suits better than
the conventional linear analysis for the free electron laser,
as the electron beam energy spread is high. 
The analysis suggests that the currently available physical parameters
would enable the generation of the coherent gamma ray of up to 100 keV.
The electron quantum diffraction suppresses the FEL action,
by which the maximum radiation energy to be generated is limited.  
\end{abstract}

\pacs{41.50.+h, 03.65.-w,52.25.Os}
\maketitle

\section{Introduction}
A coherent x-ray or gamma ray source has relevance with applications 
across various disciplines, including the condensed matter physics,
biological and medical sciences~\cite{Protein,Protein2, fast4}. 
The development of such a light source is under increased attention,
due to recent advances in the inertial confinement fusion~\cite{tabak,sonprl, sonpla}
and the free electron laser (FEL)~\cite{Free2,Fisch, sonbackward, sonlandau}.
In conventional FEL, an ultra-relativistic electron beam
interacts with the periodically laid-out strong magnets
(so-called the wiggler or the undulator)
and subsequently emits coherent photons~\cite{colson}.
While the FEL is feasible in its function, it is still not considered to
be practical as an advanced accelerator and expensive magnets are required.
There are attempts to overcome such shortcoming, 
including the effort to replace the magnets by intense lasers~\cite{Gallardo,thomson}.
Currently a laser of the visible wavelength with the intensity ranging
from $10^{19} \ \mathrm{W} / \mathrm{cm}^2$  
to $10^{21} \ \mathrm{W} / \mathrm{cm}^2$ can be generated.
There exist schemes to generate an {\it incoherent} hard x-ray or gamma
ray~\cite{thomson, gamma} using an intense laser,
based on the nonlinear Thomson scattering.
However, the difficulties in generating the {\it coherent} radiation still
persist~\cite{Gallardo,quantum}.

In this paper, we study a scheme to generate an FEL-like coherent hard x-ray or gamma
ray, where a dense relativistic electron beams of the relativistic factor $\gamma$
ranging between 10 and 1000 and the electron density between $10^{18}$ and
$10^{21}~\mathrm{cm^{-3}}$ propagates through an intense laser undulator.
While the high beam density would be an advantage in generating a coherent ray,
the high energy beam spread would be disadvantageous~\cite{wake,monoelectron, ebeam}.
We analyze the gain of the FEL via such electron beams in the framework of
a Landau damping-like theory, which is more appropriate when the beam energy spread
is high, compared to the linear analysis conventionally applied to the FEL.
Our analysis suggests that the gamma ray of up to 100 keV would be achievable.
As the wavelength of the radiated laser gets shorter, the quantum diffraction effect
becomes of more concern as it suppresses the Landau instability~\cite{sonlandau}.
The regime where the FEL scheme becomes inappropriate due to this quantum effect is identified.

The rest of the paper is organized as follows. 
The Landau damping  theory of the FEL amplification
for the dense beams and the laser undulator is introduced (Sec. II),
and the electron quantum diffraction effect on the FEL and its consequence is discussed
(Sec. III).  
The paper is concluded in Sec. IV.

\section{Landau damping theory}
\label{sec:1}
Consider an electron beam traveling in the positive $z$-direction 
and an intense laser beam propagating in the opposite direction.  
For simplicity, the laser is assumed to be linearly polarized
as $ \mathbf{E} = E_0 \cos( k_0z +ck_0t) \hat{x}$ 
and $\mathbf{B} =  -E_0 \sin( k_0z + ck_0t) \hat{y}$.
The equation of motion for an relativistic electron is
\begin{equation}
\frac{d \gamma(v) \mathbf{v}}{dt}=- e\left[ \mathbf{E} + \frac{\mathbf{v}}{c} \times  \mathbf{B} \right] \mathrm{,}
\end{equation}
where $\gamma^{-1}  = \sqrt{1-(v/c)^2} $, and $v = \mathbf{|v|}$.  
The first order perturbed motion of an electron with the initial velocity
$\mathbf{v} = v_0 \hat{z}$ is described by
\begin{eqnarray}
v^{(1)}_x &=& - 2  \frac{V_x}{\gamma_0} \sin(k_0 z + ck_0 t) \nonumber \\
v^{(1)}_y &=& v^{(1)}_z = 0  \mathrm{,} \nonumber
\end{eqnarray}
where $\gamma_0^{-2}= 1 - (v_0/c)^2$ and $V_x(E_0, k_0)  = e E_0/ m ck_0 $.
Let us now consider a co-moving electromagnetic (E\&M) wave given by $\mathbf{E} = E_g\sin(k_g x - ck_gt)\hat{x} $ and $\mathbf{B} = E_g\sin(k_g x - ck_gt)\hat{y} $.
The perturbed motion of the electron due to this E\&M wave, ignoring the non-resonance term,  is given as 
\begin{equation}
\frac{dv^{(2)}_{z}}{dt} =\frac{1}{ \gamma_0(\gamma_0^2(v_0/c)^2 + 1)} \frac{e E_g}{ m}  v^{(1)}_x  \sin(k_g x - ck_g t)
 \label{eq:3} \mathrm{.} 
\end{equation}   
By considering only the resonance term,  
\begin{equation} 
\frac{dv^{(2)}_{z}}{dt}  =  \kappa \frac{e E_g}{m_e} \sin(\phi_0 + (k_0+k_g)z + (ck_0 - c k_g ) t)  \label{eq:pendulum} \mathrm{,}
\end{equation}  
where $ \phi_0$ is the initial phase, and
\begin{equation}
\kappa = \kappa(E_0, k_0, v_0) =\frac{1}{\gamma_0^2(\gamma_0^2(v_0/c)^2 + 1)}
   \frac{V_x(E_0,k_0)}{c} \label{eq:kappa} \mathrm{.}
\end{equation}
Eq.~(\ref{eq:pendulum}) is the well-known pendulum equation for the FEL~\cite{colson}, where
the resonance condition is given as $k_g  = 2 k_0 / (c-v_0) \propto 4 \gamma_0^2 k_0 $.  
Assuming a uniform electron distribution in the $z$-direction, 
we obtain the averaged electron energy loss rate
\begin{eqnarray}
&&\frac{m}{2} \gamma_0^3 \frac{d \delta (v^{(2)}_{z})^2}{ dt} = \gamma_0^3 \kappa^2 
 \frac{e E_g^2}{ 2m} \nonumber \\
&&\times \left(  \frac{- \omega \sin(\alpha t) }{ \alpha^2} + t\cos(\alpha t) + \frac{\omega t \cos(\alpha t) }{  \alpha} \right) \mathrm{,} \label{eq:Landau} \\ \nonumber 
\end{eqnarray} 
 where $\omega =  c (k_g - k_0)$ and $\alpha = (k_0+k_g)v_0 - \omega$.
In large $t$, the second and the third term on the right hand side
of Eq.~(\ref{eq:Landau}) are averaged out,
provided that the beam density is spatially uniform. 
Equating the electron energy loss rate to the growth rate of the E\&M energy intensity,
\begin{equation} 
n_e \frac{m}{2} \gamma_0^3 \frac{d \delta (v^{(2)}_{z})^2}{ dt}= 
\frac{ d (E_g^2 /4 \pi)}{dt} = \gamma^i\frac{E_g^2 }{4 \pi} \mathrm{,}
\end{equation}
leads to the growth rate of a hard x-ray or gamma ray due to the Landau instability 
\begin{equation}
\gamma^i =  \frac{\pi}{2} \gamma_0  \frac{ \kappa^2 \omega_{\mathrm{bpe}}^2}{ q^2} \left(\frac{\partial f_e}{\partial v}\right)_{\omega_g / q} \label{eq:growth} \omega_g \mathrm{,}
\end{equation}
where $q = k_0 + k_g \cong (4\gamma_0^2 +1) k_0 $, $\omega_g$ is the frequency of the radiated field,  
$\omega^2_{\mathrm{bpe}} = 4 \pi n_b e^2 / m_e$ 
is the Langmuir frequency of the beam density, and
$f_e $ is the electron distribution function with the normalization condition
$\int f_e dv_z = 1$.
The growth rate given in Eq.~(\ref{eq:growth})
is more appropriate than the conventional FEL linear analysis~\cite{colson} in the regime where $\delta \alpha t \cong k_g \delta v t \gg 1$, which is the case for the electron beam with the high energy spread as we consider. 
Denoting the laser duration by $\tau$,
the growth factor is $G = \exp(\gamma_i \tau)$ and the necessary condition for the FEL is $\gamma_i \tau > 1$. 

Consider $1\mu \mathrm{m}$ Nd:YAG laser of the intensity $I$.  
Let us define $I_{18} = I/ (10^{18} \mathrm{W} / \mathrm{cm}^2$) and 
the energy spread  of the electron beam as $\delta E / E = \zeta $. 
Then, in a crude approximation, 
$\delta v / c \cong  \zeta / \gamma_0^2$ and  
  $(\partial f / \partial v)_{\omega/q}  \cong (1/ c^2) (\gamma_0^4/\zeta^2)$,
when $\gamma_0 \gg 1 $,   $\omega \cong (4\gamma_0^2 -1) ck_0$ 
and $q \cong (4\gamma_0^2 +1) k_0$. 
Assuming $v_0 \cong c$, $\kappa$ can be estimated as
$ \kappa^2 = (1/2.5 \gamma_0^4) I_{18}$ and Eq.~(\ref{eq:growth}) becomes 
\begin{equation}
 \gamma^i \tau  = 1.4 \times 10^{-2} \frac{n_{20} I_{18}}{\gamma_0^3 \zeta^2} \omega_0\tau   \mathrm{,}\label{eq:key}
\end{equation}
where $n_{20} =  n_b / (10^{20} \mathrm{cm^{-3}})$.
The condition $\gamma^i \tau>1$
results in the constraint between the laser intensity 
and  the electron beam energy (density and spread).  
For instance, when $\gamma_0 = 10$, 
the amplified E\&M wave has the wavelength of $2.5 \ \mathrm{nm}$ 
which should satisfy  $ (n_{20} I_{18} /\zeta^2) \omega_0\tau > 10^5$. 
If the beam energy spread is 1 \% ($\zeta = 0.01$) and $\tau= 10^{-12} \sec$,  
the beam parameters of $n_{20} \cong 0.1 $ and  $I_{18} \cong 1$ 
are sufficient to satisfy  $\gamma^i \tau > 1$.
For $\gamma_0 = 100$, where the radiation wavelength is $0.025 \ \mathrm{nm}$, 
the condition ($\gamma^i \tau > 1$)  might be still achieved by 
decreasing the energy spread and increasing the laser intensity or laser duration. 
When $\gamma_0=1000$, it becomes very difficult unless an extremely intense laser
of a long duration is available and the beam energy spread is extremely narrow.  

\section{Quantum diffraction and the Suppression of the FEL}
\label{sec:2}
The wavelength of the wiggler $\lambda_0$ in the conventional FEL is of the order of cm. 
In the Lorentz frame where the electrons are stationary, 
the wavelength $\lambda_0$ is reduced to be $\lambda_m = \lambda_0/\gamma_0$,
which is still large enough for the electron quantum diffraction to be negligible. 
However, when $\lambda_0 = 1\ \mu \mathrm{m}$, which is comparable to the wavelength of 
the Nd:YAG laser, the electron quantum diffraction effect becomes relevant. 
Previous study shows that the quantum diffraction effect suppresses the Langmuir
wave damping~\cite{sonlandau}.  
In the following, a similar analysis given in Ref.~\cite{sonlandau}, for the FEL, is provided.

In this reference frame where the electron beam is nearly stationary,
the laser frequency would be shifted up by $\omega_m = 2 \gamma_0 \omega_0 $ and 
the electric field would be increased by $E_m = 2\gamma_0 E_0$, 
where $\omega_0$ ($E_0$) is the laser frequency 
(the electric field) in the laboratory frame and $\omega_m$ ($E_m$) is
in the moving frame. 
Let us re-interpret the FEL mechanism in the moving frame. 
The laser of $\omega = \omega_m$ and $k_z =- \omega_m/c$ 
would travel from the right to the left and make the electrons oscillate,
resulting in the E\&M wave of $\omega = \omega_m $ and  $k_m = \omega_m/c$ unstable.
The unstable wave, which is  seen in the laboratory frame as 
 $\omega_g = 2 \gamma_0 \omega_m = 4 \gamma_0^2 \omega_0$, 
 gets amplified propagating from the left to the right. 
The electrons act  as the momentum reservoir through which  the laser would
give away the energy to the amplified wave~\cite{quantum}.  

For the quantum-mechanical analysis of the FEL action,
we write the Schroedinger equation in the moving frame
$H_0 \psi = i (\hbar \partial \psi / \partial t)$ as
\begin{equation} 
   H_0  = 
  \left[  \frac{p_y^2+p_z^2}{2m_e} + \frac{(p_x - eA_x/c)^2}{2 m_e} \right] \psi  \mathrm{,} \label{eq:quantum}
\end{equation}
 where $p_x = i\hbar \nabla_x $, $p_y =  i\hbar \nabla_y $, $p_z =  i\hbar \nabla_z $,  $A_x = (c/\omega_m ) E_m \cos(k_m z + \omega_m t)$,  and $k_m= 2 \gamma_0 k_0 $ ($\omega_m = 2 \gamma_0 \omega_0$) is the laser wave vector (frequency). 
The solution to the above equation is the Volkov state,
which also can be obtained from the full Dirac equation~\cite{volkov}.   
The electron quantum diffraction is relevant
when the de Broglie time scale of the electron kinetic energy is 
faster than the time scale the FEL occurs, or $\hbar k_m^2 /m_e > 1/T$, 
where the $T = \tau / \gamma_0$ is the laser duration in the co-moving frame.
For the YAG laser of $\lambda_0 = 1 \ \mu \mathrm{m}$, 
the condition is 
 \begin{equation}
  \gamma_0 >  54 / T_{-12} \mathrm{,} \label{eq:q}
\end{equation}
where $T_{-12} = T / (10^{-12} \ \sec) $.
Consider the FEL radiation given as 
$ \delta A_x =  (c/\omega_m) \delta E_{m} \cos(k_b z - \omega_m t) $, 
where $\delta E_{m}$ is the electric field of the amplified E\&M wave.  
The full Hamiltonian is given as $H = H_0 + H_1$, where $H_1 $ is
\begin{equation}
H_1 = \frac{1}{2m_e} \left[ \left(p_x - \frac{e A_x}{c}\right)  \frac{e \delta  A_x}{c}  + 
    \frac{e \delta  A_x}{c}  \left(p_x - \frac{e A_x}{c}\right) \right] \mathrm{.} 
\end{equation} 
The dominant term is the quiver term, as the laser is very intense.
$H_1$ can be approximated  as 
\begin{equation} 
  H_1 = - m_e \left( \frac{ e E_m }{m_e \omega_m }\right)\left( \frac{e \delta E_m}{ m_e \omega_m} \right)\cos(2 k_m z) \mathrm{,}
\end{equation}
where $H_1$ forms a pseudo-lattice.  
Quantum mechanically,  the FEL action could be described as 
electrons of the initial momentum $\hbar k_m$ absorb the momentum $-2\hbar k_m$  from  $H_1$.
Simultaneously, the laser photon of  $k = -k_m $  
is channeled to a photon of $k = +k_m$ due to the momentum conservation.  

If the reflected light is strong enough and $\gamma_0$ is large enough as in Eq.~(\ref{eq:q}),
the distorted electron dispersion would suppress the Fermi's golden rule,
consequently the FEL amplification.   
As was analyzed previously in Ref.~\cite{sonlandau}, the band gap size is given as
$\delta \omega_g \cong  (m_e/ \hbar) (e E_m / m_e \omega_m )( e \delta E_m / m_e \omega_m)$.  
If $\delta \omega_g T  > 1$, 
the electron momentum of the initial value $k_z = k_m$ 
would oscillate between $k$ and $-k$, instead of transitioning from $k$ to $-k$.  
This oscillation is detrimental to the FEL, as the Landau damping,
which is the energy channel for the FEL, is suppressed. 
The condition that the band gap would not entirely suppress the FEL
is given as  $\delta \omega_g < 1/T $. 
Noting that  $ T =  \tau / \gamma_0 $, it can be re-casted as  
\begin{equation}
\delta v_{\mathrm{osc}} < \frac{1}{m v_{\mathrm{osc}}} \frac{\gamma_0 \hbar }{ \tau} \mathrm{,} \label{eq:critical} 
\end{equation}
where  $\delta v_{\mathrm{osc}} = e \delta E_m / m_e \omega_m$ and 
$v_{\mathrm{osc}} = e E_m / m_e \omega_m  =  e E_0 / 2 m_e \omega_0$,
using $ \omega_m = 2 \gamma_0\omega_0$ and $ E_m = 2\gamma_0 E_0$. 
On the other hand, 
note that $\delta v_{\mathrm{osc}} =  e \delta E_L / 4m_e \omega_0 \gamma_0^2$ 
from the relationship (from the Lorentz transform), $\delta E_L = 2 \gamma_0 \delta E_m$, where  $\delta E_L$ is 
the electric field of the amplified E\&M wave in the laboratory frame. 
Eq.~(\ref{eq:critical}) in the laboratory frame is given as  
\begin{equation} 
\frac{(\delta E_L)^2}{ 8 \pi} <  I_C =  \frac{ \gamma_0^6 \hbar^2 }{ (2 \pi)^2 \tau^2 } \frac{m \omega_0^2}{ e^2} \frac{1}{I} \mathrm{,} \label{eq:critical2}
 \end{equation}
where $I = E^2_0 / 8 \pi$ is the laser intensity in the laboratory frame. 
As the gamma ray gets amplified coherently, 
the beating potential between the laser and the gamma ray generates a pseudo-lattice. 
This pseudo-lattice, in turn, suppresses the lasing 
via the electron quantum diffraction effect, limiting the maximum energy of the radiation (Eq.~(\ref{eq:critical2})).
For the laser satisfying $\gamma \tau =1 $ in Eq.~(\ref{eq:key}), 
if $n_{20}$, $\zeta$ and  $\omega_0\tau$ are fixed,  
$I_C$ in Eq.~(\ref{eq:critical2}) is proportional to $\gamma_0^3$. 
 By noting that the duration of the radiation is reduced by $\tau_g = \tau / \gamma_0^2 $,  
the possible maximum energy that can be generated into the gamma ray ($I_c \tau_g$)
is proportional to $\gamma_0$. 
This suggests that, for a given target frequency, 
a higher $\gamma_0$ and lower laser frequency is preferred to avoid the suppression
of the lasing due to the electron quantum diffraction. 

For instance, consider $\gamma_0 = 100 $, $I_{18} =1$, $\zeta = 0.001$ and  $\tau_{-12} = 1$, and $n_{20} = 1$.
Then, Eq.~(\ref{eq:growth}) is satisfied, and the radiation wavelength would be $0.025 \ \mathrm{nm}$.  From Eq.~(\ref{eq:critical2}),  we obtain the maximum total energy radiated, $I_c \tau_g$,  is  $10^{-9}$ of the input laser energy, assuming the spot-size of the electron beam and the laser are of the same order. 

\section{Conclusion} 
\label{sec:3}
The plausibility of a hard x-ray or gamma ray laser scheme based upon
a dense electron beam and an intense laser wiggler is analyzed in 
the context of the Landau damping theory.
With the currently available intense electron beam from the wake field accelerator
or the interaction of an intense laser with the metal foil, 
a coherent hard x-ray or gamma ray of up to 100 keV could be generated.
As the wavelength of the radiation gets smaller,
the electron quantum diffraction effect becomes important. 
Our focus is on the suppression of the Landau damping due to the band gap.
The electron diffraction introduces a strong constraint on the maximum radiation
intensity achievable. 

It is assumed that the Schroedinger equation (Eq.~(\ref{eq:quantum})) is valid. 
However, this is no longer the case for high intensity $ I_{18} > 1$, 
where our estimation should be modified accordingly.  
The full quantum relativistic prediction of the electron quantum diffraction effect 
might be addressed through the band gap calculation based 
on the solution of the Volkov state solution~\cite{volkov},
which should be addressed in the future research.


\end{document}